\documentclass[a4paper]{spie}
\usepackage{amsmath,amsfonts,amssymb}
\usepackage{graphicx}
\usepackage{setspace}
\usepackage{tocloft}
 
\usepackage[colorlinks=true, allcolors=blue]{hyperref}
\usepackage{xcolor}
\usepackage[normalem]{ulem}
\usepackage{verbatim}
\usepackage{mwe}
\usepackage{subfig}
\usepackage{pgfplotstable,filecontents}
\pgfplotsset{compat=1.9}
\usepackage{geometry}

\title{CCAT: Characterization of the first science-grade MKID array for the Prime-Cam 850 GHz module}

\author[a]{Anthony I. Huber}
\author[b]{Jordan Wheeler}
\author[a]{James Burgoyne}
\author[c,a,d]{Scott Chapman}
\author[b]{Jason Austermann}
\author[b]{James Beall}
\author[e]{Steve K. Choi}
\author[d]{Doug Henke}
\author[b]{Johannes Hubmayr}
\author[f]{Ben Keller}
\author[b]{Matthew Koc}
\author[b]{Jeff van Lanen}
\author[g]{Tilak Patel}
\author[h]{Adrian Sinclair}
\author[b]{Anna Vaskuri}
\author[g,f]{Eve M. Vavagiakis}
\author[b]{Michael Vissers}
\author[a]{Ruixuan (Matt) Xie}

\affil[a]{Dept. of Physics and Astronomy, University of British Columbia, Vancouver, BC, Canada}
\affil[b]{Quantum Sensors Division, National Institute of Standards and Technology, Boulder, CO, USA}
\affil[c]{Dept. of Physics and Atmospheric Science, Dalhousie University, Halifax, NS, Canada}
\affil[d]{NRC Herzberg Astronomy \& Astrophysics Research Centre, Victoria, BC, Canada}
\affil[e]{Center for Experimental Cosmology and Instrumentation, Department of Physics and Astronomy, University of California, Riverside, CA, USA}
\affil[f]{Department of Physics, Cornell University, Ithaca, NY, USA}
\affil[g]{Department of Physics, Duke University, Durham, NC, USA}
\affil[h]{William H.\ Miller III Department of Physics and Astronomy, Johns Hopkins University, Baltimore, MD, USA}

\authorinfo{Further author information: Send correspondence to A.I.H.: E-mail: tony.huber@ubc.ca}

\cftpagenumbersoff{figure}
\cftpagenumbersoff{table}
\newgeometry{left=1.93cm,right=1.93cm,top=2.54cm,bottom=4.94cm}

\newsavebox{\measurebox}

\begin{document} 
\maketitle

\begin{abstract}
The Fred Young Submillimeter Telescope (FYST) is a 6-meter crossed-Dragone telescope developed by the CCAT collaboration.
Sited at 5600\,m on Cerro Chajnantor in the Atacama Plateau of Chile, FYST aims to provide superior atmospheric transmission and a wide field of view for submillimeter observations. 
Prime-Cam is a first-generation instrument for FYST designed to house up to seven separate instrument modules.
Among these, the 850\,GHz module represents the highest-frequency band and is optimized for ultra-sensitive broadband polarimetry and imaging. 
This module is designed to deploy $\sim$38,000 polarization-sensitive lumped-element titanium-nitride (TiN) microwave kinetic inductance detectors (MKIDs) across three arrays.
Thus the 850\,GHz module will have the most submillimeter MKIDs in a single instrument module to date.
These detectors have adopted a novel two-octave design to maximize the multiplexing achieved using a RFSoC readout system.
We review the design parameters and fabrication process for the first 850\,GHz science-grade array.
The preliminary measurements of the array show a fabrication yield of 99\%, highlighting the successful application of the design and fabrication process.
We further present the cryogenic characterization of the first full MKID array developed for the Prime-Cam 850\,GHz module.
Multiple tests were performed on the devices including resonator frequency mapping, quality factor measurements, optical load sweeps and noise performance.
From these measurements, we discuss the measured optical efficiency, sensitivity, and uniformity across the array and the expected on-sky performance of the module.
The 850\,GHz module will be deployed for observing in 2027.
\end{abstract}


\section{INTRODUCTION}

Far-infrared (FIR) and submillimeter observations are crucial for understanding dusty galaxy evolution, star formation, magnetic fields, large scale structure, and the early Universe.\cite{Farrah,CV,DS}
Roughly half of the radiative energy emitted throughout cosmic history is observed within the far-infrared and submillimeter regime due to a combination of thermal emission from cool dust and cosmological redshifting of distant sources.\cite{Lutz}
Observations in this wavelength range are therefore uniquely suited to address a broad range of questions in modern astrophysics.
However, ground-based observations in the submillimeter range are fundamentally constrained by atmospheric transmission.
At frequencies approaching 850 GHz, detector sensitivity alone is insufficient, and access to high altitude, exceptionally dry observing sites becomes critical to instrument performance.

For this reason the CCAT collaboration developed the Fred Young Submillimeter Telescope (FYST, see Fig.$~$\ref{fig:fyst_primecam} (Left)), a 6-meter crossed-Dragone telescope optimized for millimeter and submillimeter observations from Cerro Chajnantor in northern Chile at an elevation of 5600 m.\cite{P1,P2,Niemack:16}
The crossed-Dragone optical design and the Cerro Chajnantor site combine a large diffraction-limited field of view together with exceptional atmospheric transmission, particularly within the 850 GHz atmospheric window.\cite{CCAT-Collab:23}
These capabilities position FYST to deliver mapping speeds and survey depth beyond previous terrestrial and spaceborne facilities operating in comparable wavelength bands.

\subsection{Prime-Cam and the 850 GHz Module}

\begin{figure}[t]%
    \centering
    \includegraphics[width=0.45\linewidth]{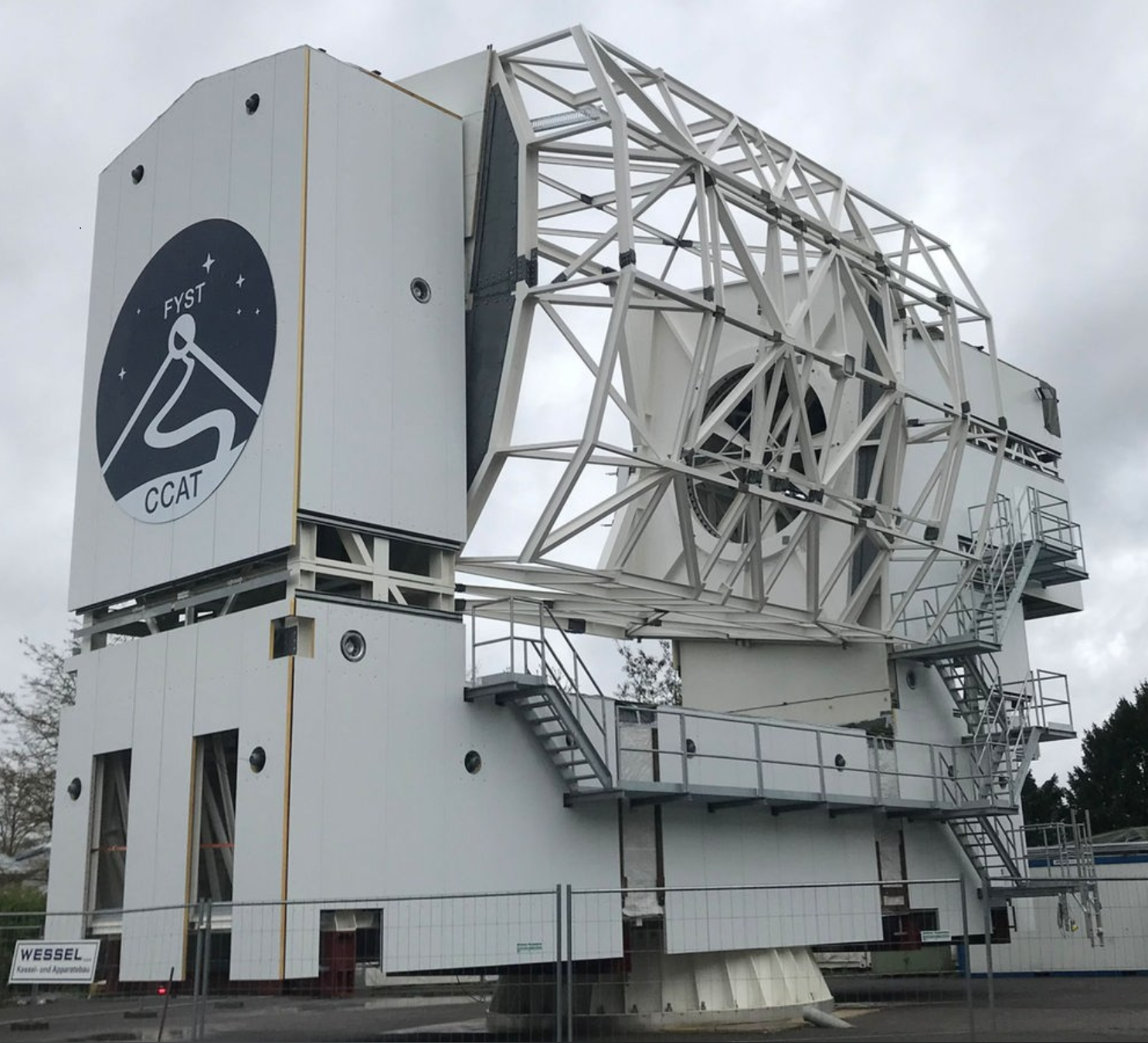}%
    \includegraphics[trim={0cm 4.2cm 1cm 5cm}, clip, width=0.45\linewidth]{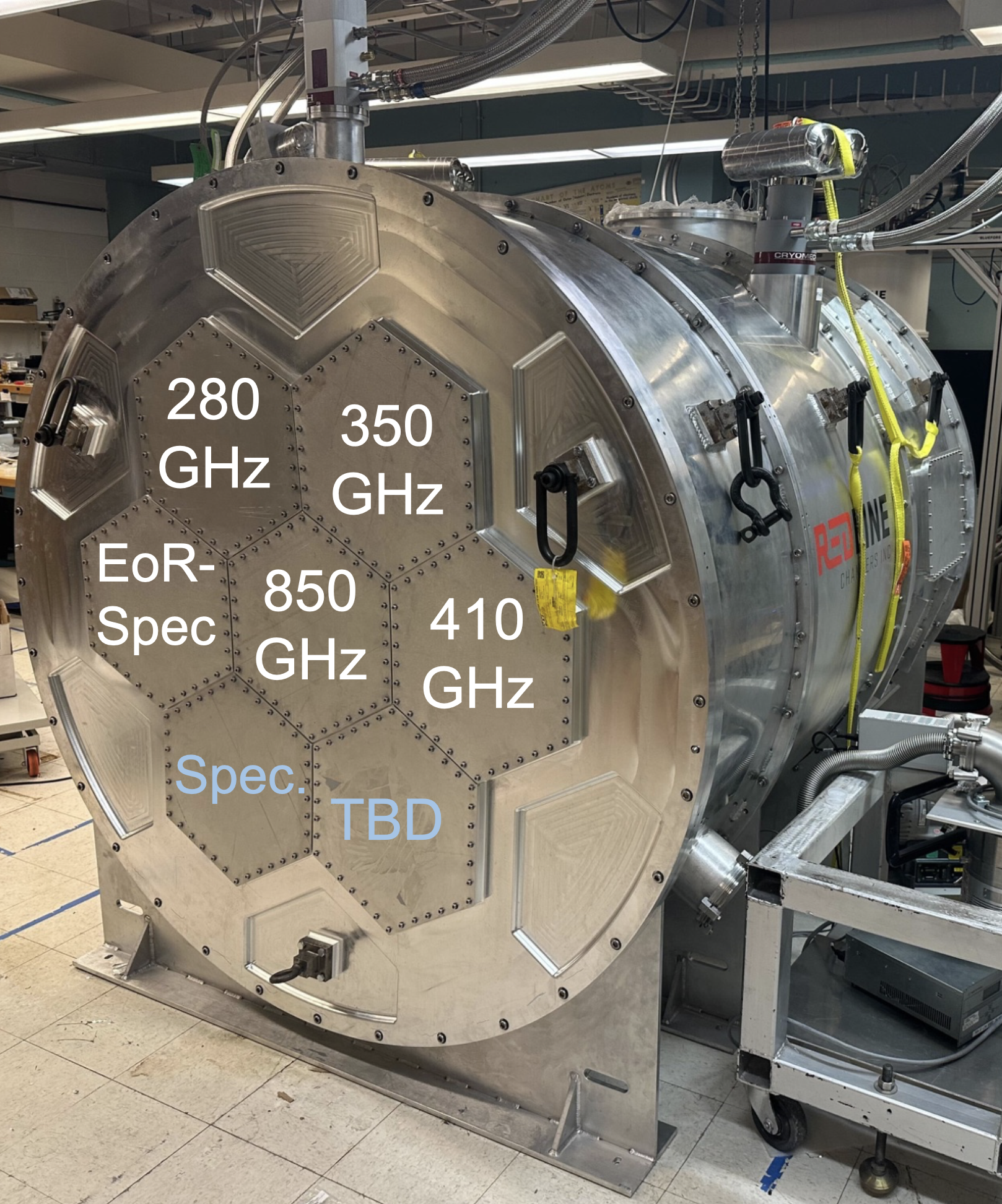}%
    \caption{Left: The Fred Young Submillimeter Telescope (FYST), a 6-meter crossed-Dragone telescope designed for operation on Cerro Chajnantor. Right: Prime-Cam instrument architecture highlighting the central placement of the 850 GHz module.}%
    \label{fig:fyst_primecam}%
\end{figure}

Prime-Cam is a first-generation instrument for FYST designed around a modular architecture in which independent instrument modules occupy distinct regions of the telescope focal plane\cite{Eve} (Fig.$~$\ref{fig:fyst_primecam} (Right)).
The baseline instrument incorporates multiple camera and spectrometer modules spanning frequencies from the millimeter to submillimeter regime, each optimized for a specific set of science goals.
Among these instruments the 850 GHz module, the highest frequency, occupies a privileged central position within the instrument architecture\cite{AHuber2022}.
The 850 GHz band is of particular importance for studies of dusty star-forming galaxies, infrared luminosity functions, and obscured star formation at high redshift.\cite{Scott2022} Observations in this band provide strong constraints on dust spectral energy distributions and complement lower-frequency measurements throughout the Prime-Cam instrument suite.
For details on the optical design of the 850 GHz module, see Ref.$~$\citenum{AHuber2022}.

The scientific requirements of the 850 GHz module demand a combination of high mapping speed, broad field coverage, and high detector density.
Achieving these goals requires both a highly optimized optical design and an unprecedented detector implementation.
The current design for the module incorporates approximately 38,000 polarization-sensitive lumped-element kinetic inductance detectors (LEKIDs or KIDs) distributed across the focal plane, representing one of the most densely packed terrestrial submillimeter detector systems developed to date.

LEKIDs provide several advantages for large-format submillimeter arrays, including relatively simple lithographic fabrication, natural frequency-domain multiplexing, and demonstrated performance across a wide range of astronomical instruments.\cite{Doyle:08}
However, deploying detector counts approaching $\sim 4\times10^4$ detectors introduces substantial challenges in readout architecture, cryogenic radio frequency (RF) design, and detector frequency planning. Conventional multiplexing strategies quickly become impractical when scaled to the required detector count.
The 850 GHz module is scheduled for deployment in 2027.

\subsection{Motivation for a Science-Grade Array}

The development path toward the deployable Prime-Cam 850 GHz focal plane has included multiple generations of detector design studies, optical prototypes, and frequency-planning investigations.\cite{Huber:24:SPIE,Huber:PhD}
Earlier development efforts established the feasibility of densely packed dual-polarization titanium nitride LEKIDs together with a two or more octave readout strategy designed to substantially increase multiplexing density.
A key innovation of the 850 GHz detector architecture is the implementation of a two-octave resonator design and the associated readout hardware\cite{Xie:26}.
The detector frequency plan combines coarse adjustment through selectively shorted inductor geometries with fine tuning through modifications to the interdigitated capacitors\cite{Huber:24:SPIE}.
This approach enables detector frequencies to be distributed across multiple bands within the expanded readout bandwidth while minimizing changes to absorber geometry and optical coupling performance\cite{Huber:24:ASC}.

The transition from prototype studies to a deployable focal plane requires characterization of a science-grade detector implementation fabricated using the finalized process stack and detector layout.
Unlike earlier development devices or dedicated witness structures, the science-grade array represents the first realization of the full detector architecture intended for instrument deployment.
The array presented in this work consists of dual-polarization titanium nitride LEKIDs fabricated at the National Institute of Standards and Technology (NIST) using a seven-layer process developed for the Prime-Cam 850 GHz detector program.
Ongoing characterization efforts focus on sampled regions of the final array and include dark measurements, optical testing, and preliminary noise studies.

In this work we first review the finalized detector architecture and fabrication approach adopted for the first science-grade 850 GHz array.
We then describe the cryogenic setup used to evaluate sampled regions of the array under both dark and optical loading conditions.
Finally, we present the characterization of the multi-octave MKIDs with the implications for detector yield, sensitivity, and projected instrument performance.

\section{DETECTOR ARCHITECTURE}

The detector architecture adopted for the Prime-Cam 850 GHz module is driven by three primary requirements: high optical efficiency at 850 GHz, aggressive focal plane packing density, and sufficiently large multiplexing factors to maintain practical cryogenic and readout complexity.
Meeting these requirements simultaneously demands careful optimization of detector layout, resonator design, and frequency planning.
The 850 GHz focal plane utilizes feedhorn-coupled dual-polarization titanium nitride lumped-element kinetic inductance detectors (LEKIDs).
Each spatial pixel incorporates two orthogonal LEKIDs to provide polarization sensitivity while maintaining compatibility with the optical coupling scheme of the instrument.

\subsection{Dual-Polarization TiN LEKID Design}

\begin{figure}[t]
\centering
\includegraphics[width=0.7\linewidth]{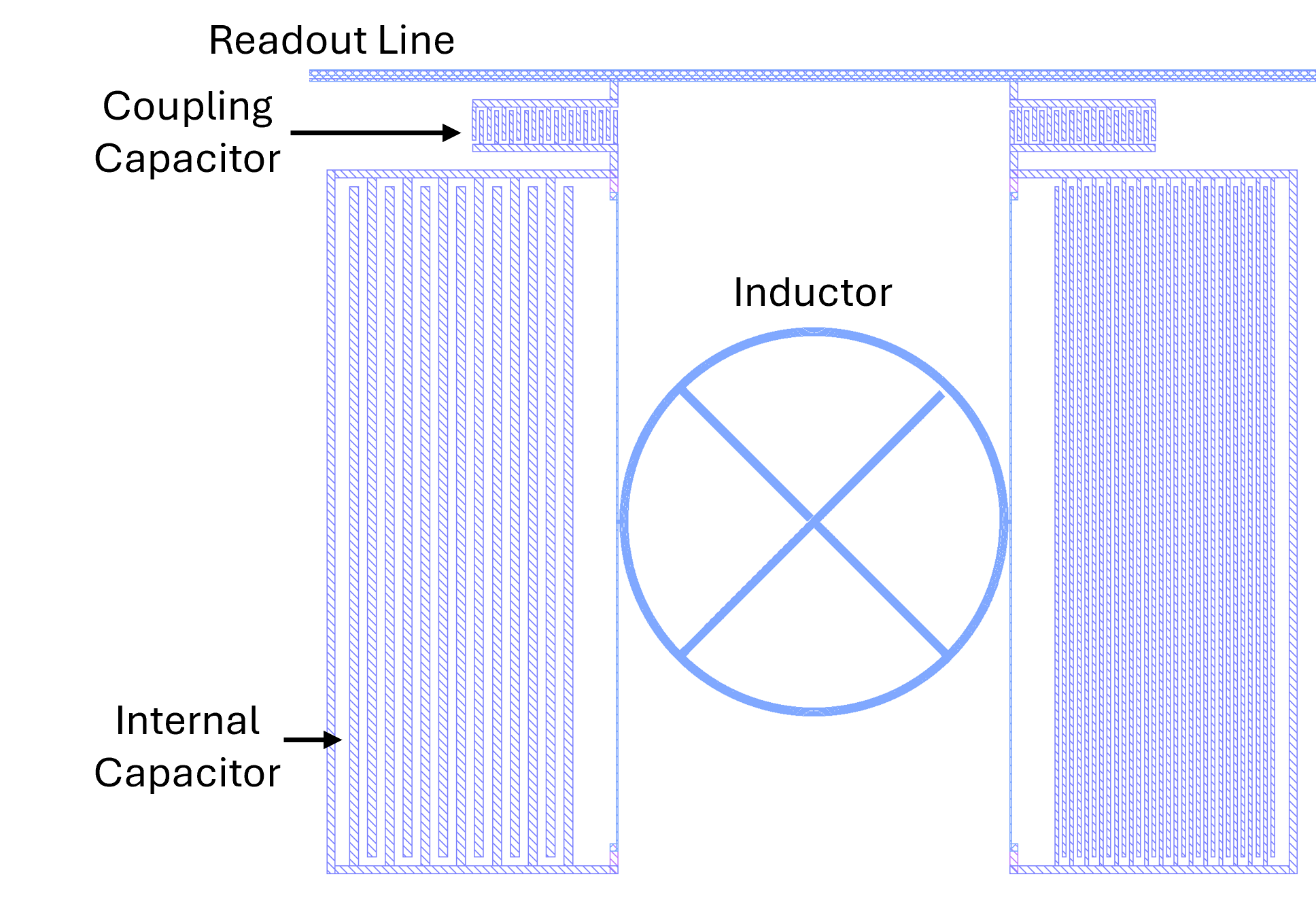}%
\caption{
Rendering of a single 850 GHz pixel consisting of two orthogonal lumped-element kinetic inductance detectors including labels for discrete components. Image from Ref.$~$\citenum{Huber:24:SPIE}.
}
\label{fig:pixel_layout}
\end{figure}

The detectors presented in this work employ a lumped-element architecture in which the resonant inductive and capacitive elements are physically distinct (see Fig.$~$\ref{fig:pixel_layout}).
The inductive element functions simultaneously as the resonator inductance and optical absorber, while the resonant frequency is defined primarily through the combination of the inductor and interdigital capacitor (IDC).
This intrinsic resonance allows MKIDs to be multiplexed with hundreds of MKIDs in a single line or network.
Titanium nitride (TiN) was selected as the detector material due to its suitability for high-frequency submillimeter absorber implementations, tunable superconducting properties, and compatibility with dense detector geometries.
In contrast to lower-frequency implementations with larger pixel pitches\cite{Choi:22,Keller:26,Freundt:24,Patel:26}, the 850 GHz module requires detector footprints compatible with extremely compact focal plane packing.

Dense detector packing strongly constrains the allowable resonator footprint.
As a result, the detector geometry represents a compromise between maximizing detector performance and minimizing the total pixel area.
The resulting design enables the deployment of approximately 38,000 detectors across the planned focal plane while maintaining compatibility with the optical requirements of the module.
Each pixel contains a pair of detectors with orthogonally oriented absorbers to enable dual-polarization sensitivity.
The absorber geometries are optimized to preserve the polarization response while satisfying competing constraints arising from detector volume, optical coupling, resonator frequency placement, and fabrication tolerances.

\subsection{Two-Octave Frequency Planning}

One of the dominant technical challenges for the 850 GHz module is multiplexing a detector count approaching $4\times10^4$ detectors within practical cryogenic RF constraints.
Traditional KID readout implementations typically accommodate approximately 500 detectors/network which, when scaled to the required 850 GHz focal plane size, would demand a prohibitive number of cryogenic RF chains, low noise amplifiers, coaxial lines, and readout electronics.
The 850 GHz program therefore incorporates a multi-octave frequency strategy designed to significantly increase multiplexing density per readout channel.

The 850 GHz detector array utilizes a two-octave resonator architecture based on a combination of coarse and fine frequency tuning techniques.
Coarse tuning is achieved through modifications to the inductance by shorting pairs of inductor lines.
These shorts reduce the effective inductance of the resonator and permit detectors to be distributed across multiple frequency groupings while preserving a common absorber framework.
Fine tuning is achieved through adjustments to the interdigital capacitors. Variations in IDC geometry provide localized control of resonant frequency placement within each coarse tuning band and allow implementation of the final frequency plan.

This approach substantially expands the accessible resonator frequency space while minimizing disruption to detector optical behavior.
Since the absorber geometry remains largely unchanged across all pixels, the method preserves consistency in detector optical performance while enabling increased multiplexing density.
For full details on the two-octave architecture, see Refs.$~$\citenum{Huber:24:ASC,Huber:24:SPIE}

\subsection{Readout Considerations}

The two-octave strategy is designed for compatibility with Radio Frequency System on a Chip (RFSoC) based readout hardware, gateware, and software developed for large-scale detector multiplexing\cite{Adrian,Sinclair:24,Xie:24}.
Scaling to the required detector count introduces challenges extending beyond detector fabrication alone. Large detector populations require careful management of tone placement, resonator collisions, amplifier dynamic range, calibration procedures, and data throughput.
These considerations influence acceptable resonator spacing and frequency planning requirements within the science-grade array.
The two-octave implementation reduces the required number of cryogenic RF channels and associated hardware overhead.
Each array will be split into 12 networks with more than 1,000 detectors per network.
In parallel, the RFSoC platform provides a flexible environment for generating, tracking, and processing large resonator tone populations.
A detailed description of the two-octave RFSoC readout architecture and preliminary performance is presented in Ref.$~$\citenum{Xie:26}.

\section{EXPERIMENTAL SETUP}
\label{sec:setup}
\begin{figure}[p]
    \centering
    \includegraphics[trim={0cm 2cm 0cm 0cm}, clip, width=0.85\linewidth]{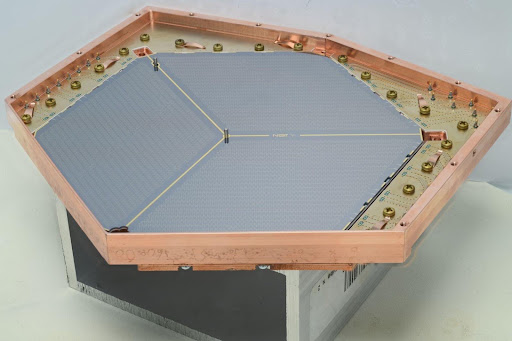}%
    \caption{The first science-grade detector array for the Prime-Cam 850 GHz module. The array was fabricated by NIST using stacked layers of TiN and Ti, and incorporated the two-octave design presented in Refs.$~$\citenum{Huber:24:SPIE,Huber:24:ASC,Huber:PhD}.}
    \label{fig:Detectors}

    \vspace{\floatsep}

    \centering
    \includegraphics[trim={20cm 50cm 0cm 35cm}, clip, width=0.8\linewidth]{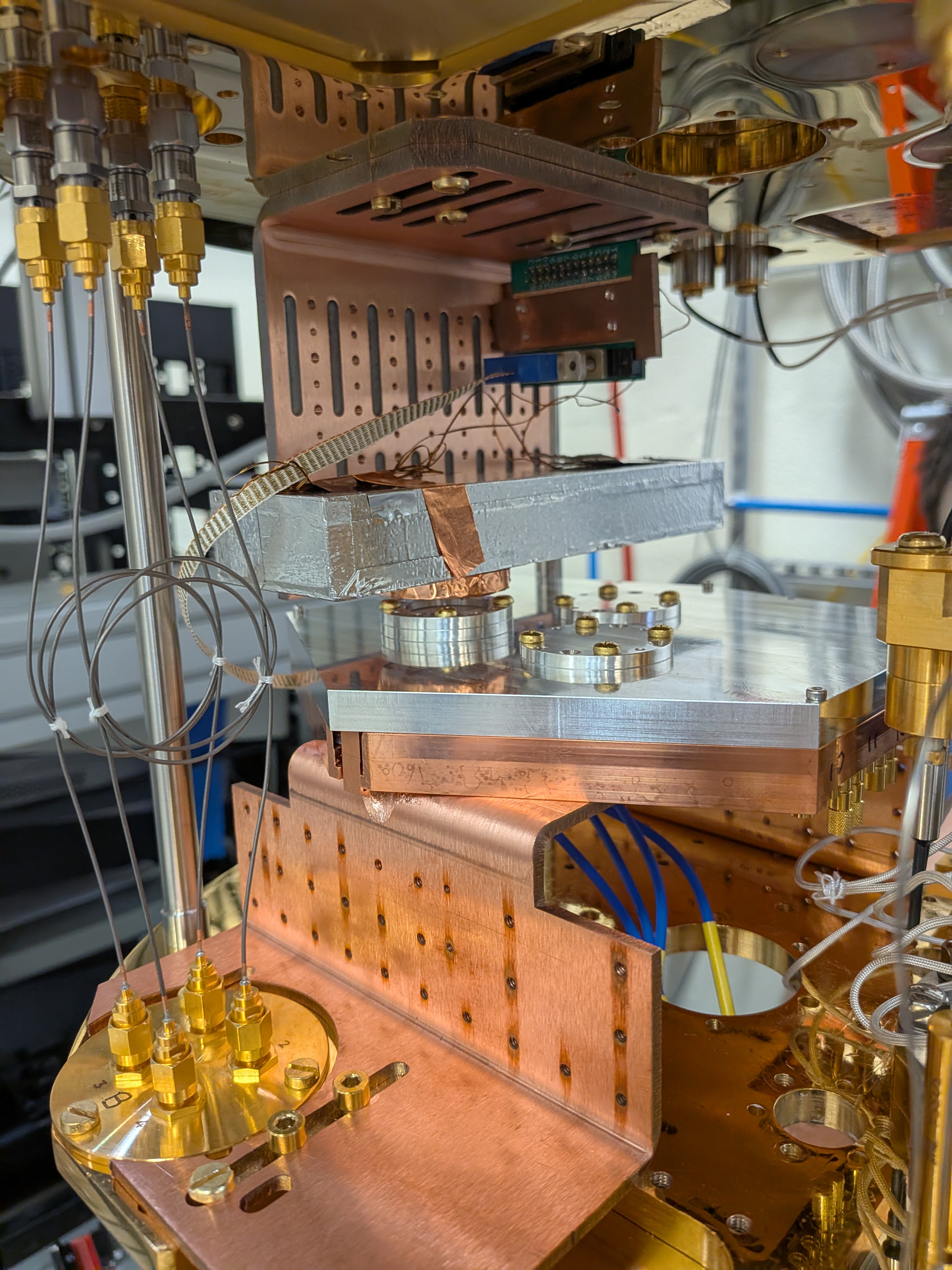}%
    \caption{Experimental setup of the blackbody (top) relative to the test array (bottom). Due to size constraints, the system is designed to accommodate three different sample regions across the array to be tested individually. The 850 GHz in a testing configuration with the blackbody and filters defines the incident power on a given sample region of the array.}
    \label{fig:Setup}

\end{figure}

The detector array characterized in this work, shown in Fig.$~$\ref{fig:Detectors}, represents the first science-grade realization of the Prime-Cam 850 GHz detector architecture.
Unlike earlier prototype devices intended to explore isolated design questions, the present implementation adopts the finalized detector framework, resonant frequency tuning strategy, and fabrication approach intended to inform instrument deployment.
Fabrication was carried out at NIST on a 6-in (152.4-mm) wafer using a seven layer TiN/Ti/TiN multilayer deposition process developed for the Prime-Cam 850~GHz detector program\cite{Vissers:13}.

Characterization of the first science-grade detector array was carried out through a combination of dark and optical cryogenic measurements.
The experimental campaign is intended to evaluate detector resonance behavior, responsivity, noise performance, and preliminary array uniformity.

Measurements were performed in a cryostat configured for operation of submillimeter MKIDs.
To reduce systematic uncertainties in the results, the test setup is held in a stable temperature-controlled environment and read using a characterized RF infrastructure at NIST.
In the test configuration, the detector package, readout chain, and optical loading configuration were selected to maximize the characterization of the MKID architectures across each of the channels.
A small series of regions were sampled due to limitations in the size of the calibrated cryogenic blackbody and corresponding filters for the array (see Fig.$~$\ref{fig:Setup}).
These regions were selected to provide a representative sample of the detectors across as many networks as possible.

Optical characterization is intended to evaluate the detector response under controlled loading conditions representative of the intended operating environment.
Measurements include optical response studies together with a preliminary assessment of detector sensitivity, loading behavior, and detector-to-detector consistency across sampled regions of the science-grade array.
As optical characterization is ongoing, this work emphasizes the measurement framework together with early results.   

\section{CHARACTERIZATION RESULTS}

Characterization of the first science-grade Prime-Cam 850 GHz detector array remains ongoing at the time of writing.
The results presented here focus on sampled regions (see Fig.$~$\ref{fig:Setup}) extracted from the final detector implementation and are intended to establish the measurement framework together with early performance indicators.
The present analysis includes resonator identification and preliminary optical response studies.
Metrics discussed below should be considered preliminary and subject to refinement as characterization progresses.

\subsection{Resonator Identification and Frequency Placement}

\begin{figure}[t]
\centering
\includegraphics[trim={4.2cm 0.5cm 5cm 0cm}, clip, width=\linewidth]{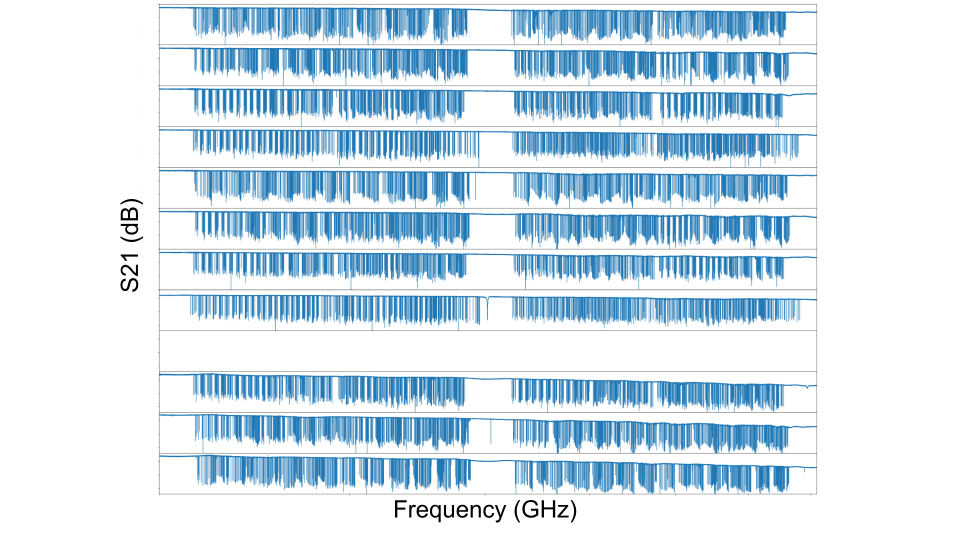}%
\caption{
First results depicting the S$_{21}$ measurements of all 12 networks on the test-array. The gap was intentionally included to create greater separation between the non-shorted (left) and shorted (right) architectures. Inspection of the ninth network found a small scratch across the feedline. However, even with the loss of a whole network the array achieved greater than 90\% fabrication yield.}
\label{fig:VNA_Sweeps}
\end{figure}

The initial test performed was a sweep measuring the complex forward transmission, S$_{21}$, across each of the networks, the first glimpse of the results of fabricated LEKIDs.
The result is shown in Fig.$~$\ref{fig:VNA_Sweeps}.
Our designs succeeded in yielding more than 1,000 detectors across two octaves.
From Fig.$~$\ref{fig:VNA_Sweeps}, there is clear delineation between the shorted and non-shorted inductor lines in each of the sweeps seen as a small gap near the center of the band.
Inspection of the wafer showed a small scratch on the feedline of the ninth network.
However, even with the loss of a whole network a greater than 90\% fabrication yield was achieved.

Following the initial S$_{21}$ measurements, the array underwent an LED mapping procedure to map the physical positions of the resonators on each array to their resonant frequencies\cite{Middleton:25,Burgoyne:26}.
Knowing the physical positions of the resonators not only enables capacitor trimming to reduce collisions in the readout, but also allows us to identify the exact responses of each detector relative to surrounding resonators.
This is particularly important in ensuring that each of the detectors is beam-filled in the optical characterization of the array.
Full details of the LED mapping procedure can be found in Ref.$~$\citenum{Burgoyne:26}.

The second objective of these preliminary measurements is verification of the two-octave frequency architecture in the detector array implementation.
While initial measurements showed there is a measureable deviation in the fractional frequency offset between the shorted and non-shorted architecture\cite{Burgoyne:26}, there is no discernible difference between the performance of the resonators themselves.
In addition to validating the detector tuning framework, these measurements provide important feedback regarding fabrication uniformity and process stability within the implementation of the detector design.

\subsection{Optical Characterization}

Once the networks were confirmed to be operational and the resonators mapped, optical characterization could commence using the setup detailed in Sec.$~$\ref{sec:setup}.
By controlling the blackbody temperature and knowing the passband of the filters, we were able to incrementally increase the power incident on the detectors until we reached our expected loading level, which for 850 GHz is 70 pW on our best observing days\cite{CCAT-Collab:23}.
From the LED mapping analysis we are able to map which resonators were responding to variations in the source. 
This allows us to not only verify the position of each detector, but also determine whether the detectors are well centered on the blackbody source, so we can begin to characterize the differences, if any, we see between the detector types.

\begin{figure}[t]
\centering
  \subfloat{\includegraphics[width=0.50\textwidth]{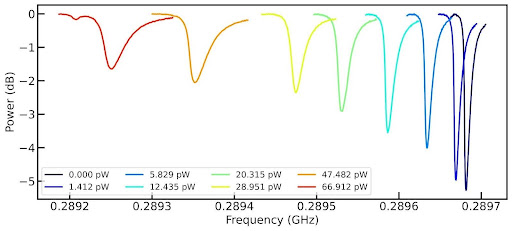}}
  \hfill
  \subfloat{\includegraphics[width=0.50\textwidth]{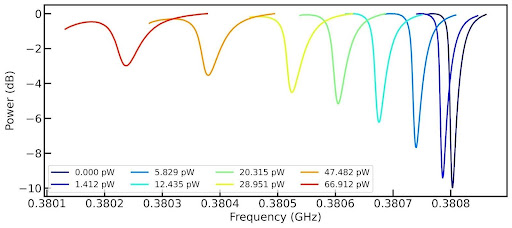}}
  \vspace{0.01em} 
  \subfloat{\includegraphics[width=0.50\textwidth]{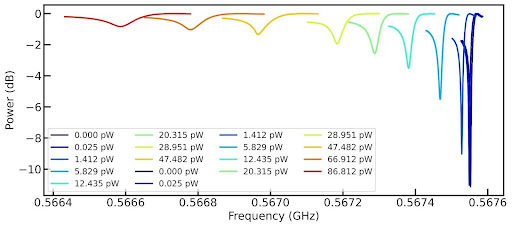}}
  \hfill
  \subfloat{\includegraphics[width=0.50\textwidth]{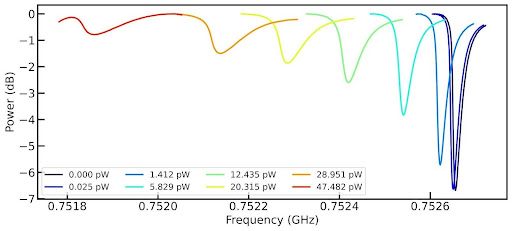}}
\caption{Resonator response to increasing optical loading. Each panel depicts the response for representative resonator within a given band based on the four detector architectures. The non-shorted resonators are on top, while the bottom panels present shorted resonators. While the frequencies shift substantially, the resonance depths, linewidths, and fit quality across all the designs remain similar.}
\label{fig:four_images}
\end{figure}

\begin{figure}[p]
    \centering
    \begin{minipage}[t]{0.4\textwidth}
        \vspace{0pt}
        \centering
        \includegraphics[width=\linewidth,height=3.8cm]{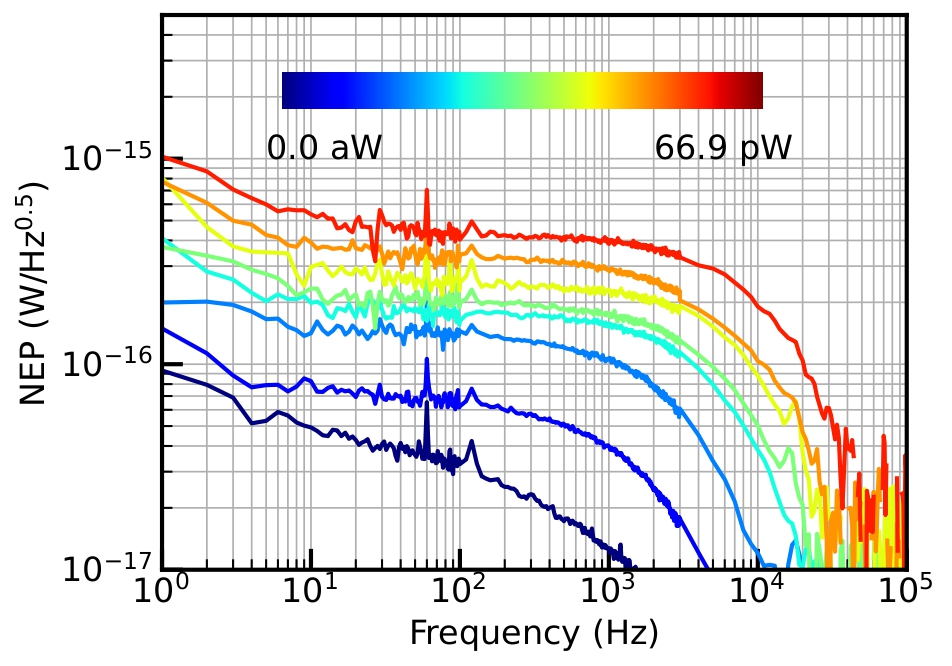}
        \includegraphics[width=\linewidth,height=3.8cm]{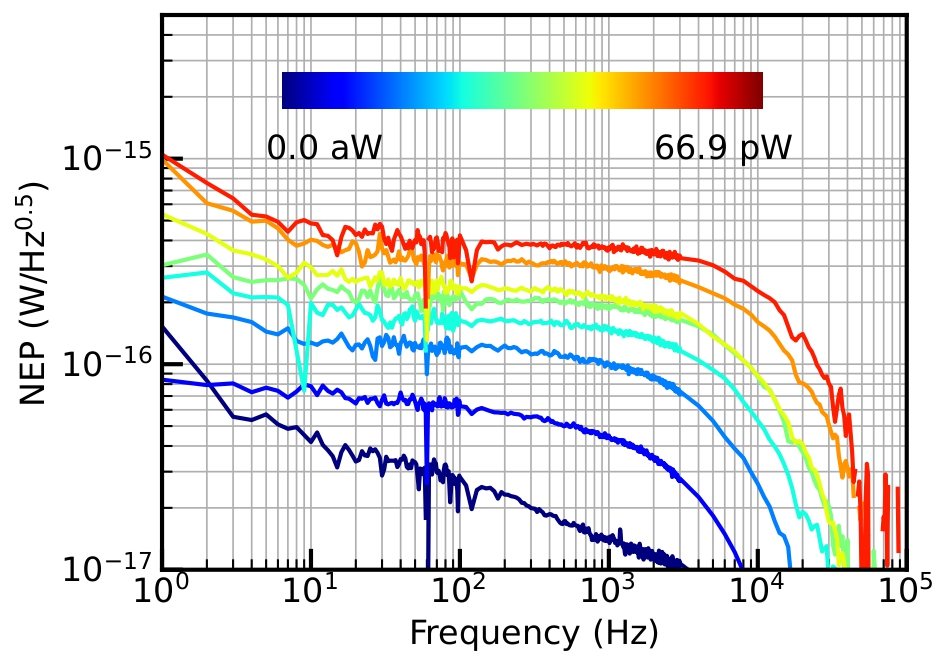}
    \end{minipage}
    \hfill
    \begin{minipage}[t]{0.58\textwidth}
        \vspace{0pt}
        \centering
        \includegraphics[width=\linewidth,height=7.6cm]{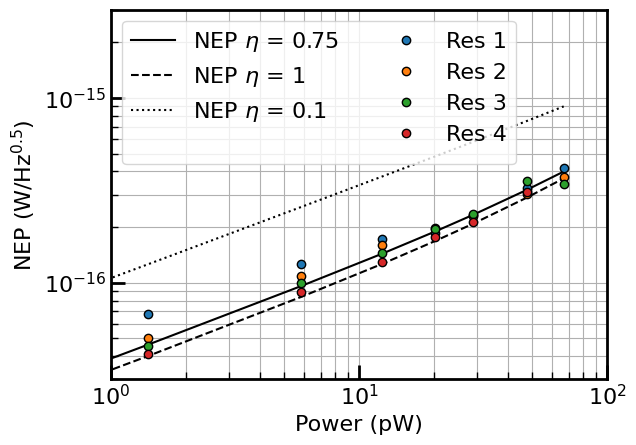}
    \end{minipage}
    \caption{Left: measured NEP values for shorted (top) and non-shorted (bottom) MKIDs. Right: Extracted white NEP for four representative resonators. Values matched the modelled optical efficiency of 75\% under loading.}
    \label{fig:NEP}

    \vspace{\floatsep}

    \centering
    \includegraphics[trim={0cm 0cm 0cm 0.9cm}, clip,width=0.75\linewidth]{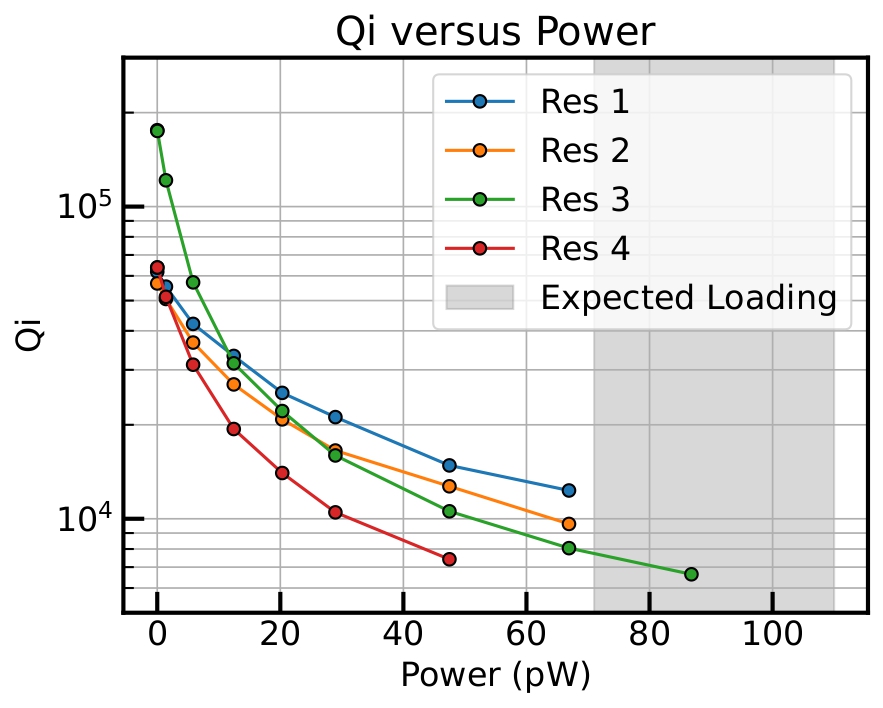}%
    \caption{Internal quality factor for four representative MKIDs under increasing optical loading. Fits show values for Q$_i$ converge within expected detector-to-detector variations under expected loading.}
    \label{fig:Qi_v_P}
\end{figure}

We began by performing S$_{21}$ sweeps across the arrays, verifying the optical response of the resonators.
Figure$~$\ref{fig:four_images} shows the shift observed in four representative resonators for the two capacitor sizes and inductor types.
As expected, the resonance frequencies shift substantially, but the resonance shapes remain similar between the resonators. 
Analysis shows comparable resonance depths, linewidths, and fit quality across each of the designs.
This is a preliminary qualitative indication that the two-octave architecture modifies resonance frequency without introducing significant systematic differences in the MKID performance; from a detector perspective, these designs are responding within observed detector-to-detector variations.
This was further confirmed in measurements of the fractional frequency shift of the sampled portion of the array, which showed minimal deviations between the detectors.

The left panel of Fig.$~$\ref{fig:NEP} presents the measured noise equivalent power (NEP) values for two MKIDs, shorted and non-shorted.
Despite occupying different frequency bands in the readout, the detectors exhibit essentially the same sensitivity.
Additionally, even though the capacitors are relatively small by design, the NEP shows that the two-level system (TLS) noise\cite{Noroozian:09} is subdominant to the photon noise, easily seen as the difference between the loaded and unloaded response at low frequencies.
The right panel shows the extracted white NEP of four representative resonators.
Preliminary results show the measured white NEP matches the expected modelled optical efficiency of the array of 75\% under loading.
Note that due to differences in the filtering of the experimental setup and the actual module, the expected optical efficiency of the deployment band is projected to be $>$90\%.
This result confirms both the detectors and feedhorns work as expected.

To further probe differences in the response between the varying architectures, differences in the quality factor were examined under loading based on existing models\cite{Gao,Mauskopf:18}.
Figure$~$\ref{fig:Qi_v_P} depicts the internal quality factor, Q$_i$, as a function of optical power for each of the four detector designs.
If the architectural modifications were significantly impacting the absorber, quasiparticle generation, or loss mechanisms we would expect the curves to separate under loading.
Observations instead show that the four MKIDs follow essentially the same trend with loading, with the remaining differences being at a level comparable to typical detector-to-detector variations.
This further supports the idea that the shorted and non-shorted architectural variations are largely decoupled from the detector response itself.

\section{DISCUSSION}

The science-grade detector array presented here represents an important transition in the development pathway of the Prime-Cam 850 GHz module.
As detector testing remains ongoing, metrics presented here should be interpreted as preliminary.
Previous work established the feasibility of densely packed dual-polarization TiN LEKIDs together with a two-octave frequency strategy designed to substantially increase multiplexing density.\cite{Huber:24:SPIE,Huber:24:ASC,Huber:PhD}.
Practical deployment depends strongly on achieving increased multiplexing density without unacceptable degradation in detector performance.
The present work extends those efforts into characterization of a science-grade implementation fabricated using the finalized detector architecture.
Of particular importance is confirmation that the two-octave detector architecture has a minimal impact on the detector performance.
Successful operation of the two-octave frequency framework is central to the overall detector strategy, and the detector count required by the 850 GHz focal plane places substantial pressure on conventional readout architectures.
Verification of the combined coarse- and fine-tuning implementation therefore represents a key milestone for the program.

The optical characterization setup has been designed to provide a direct comparison between measured detector behavior and the loading conditions expected for operation on Cerro Chajnantor.
Characterization of sampled regions from the final detector array provides insight into fabrication reproducibility within the science-grade implementation.
From the results detailed above, evidence is convincing that the two-octave architecture is in fact decoupled from the overall response of the detector.
This is observed in both the frequency response and Q$_i$ under loading.
Each surveyed resonator displayed a similar photon noise limited sensitivity under loading, and the optical efficiency was confirmed to match the expected value of 75\% for the test setup.
As a whole, this lends credence to the claim that a shorted inductor architecture, coupled with a suitable RFSoC readout, can provide a simple means of utilizing substantially more frequency space without paying a noticeable penalty in detector performance.

However, the results above also show a trend in which resonators perform suboptimally under expected loading conditions.
In particular, Fig.$~$\ref{fig:Qi_v_P} shows that Q$_i$ drops below the desired value of 15,000 even before 70$~$pW of optical loading.
Early measurements have also concluded that the coupling quality factor, Q$_c$, is 1.5 times higher than expected.
The impact of these factors is clearly visible in Fig.$~$\ref{fig:four_images}: under loading, the MKIDs enter a regime where the resonators are shallow.
As a result, under loading the readout becomes more susceptible to amplifier noise.
To combat this, corrections will be made via post-fabrication trimming, and improvements will be made on future arrays.

The characterization effort continues, and several analyses described here will continue to evolve as additional measurements become available.
Future work will expand the testing across larger regions of the detector array, refine optical calibration, and further evaluate the detector behavior within the intended readout framework.


\section{CONCLUSIONS}

We have presented the preliminary characterization campaign of a science-grade detector array developed for the Prime-Cam 850 GHz module.
The detector architecture utilizes densely packed dual-polarization TiN LEKIDs fabricated by NIST.
The design incorporates a two-octave frequency planning strategy based on coarse tuning through shorted inductor geometries together with fine tuning through variations in the interdigital capacitors.
Using this method, more than 10,000 submillimeter MKIDs were successfully produced on a single silicon wafer, the most to date.
This approach is intended to support the aggressive multiplexing requirements of the approximately 38,000-detector focal plane while maintaining compatibility with RFSoC-based readout.
Preliminary results confirm that the inductor architecture is decoupled from the response of the resonator.
Furthermore, this work confirms the expected optical efficiency and photon noise-limited performance of the array.

The present work represents an important transition from earlier detector development efforts toward the characterization of a deployment-oriented detector implementation.
The results obtained from the science-grade array will inform continued optimization of detector performance, readout integration, and projected instrument sensitivity.
Characterization efforts are ongoing, and future work will expand analysis across additional regions of the array while refining optical calibration, detector performance estimates, and deployment projections.

The Prime-Cam 850 GHz module is being prepared for deployment in 2027.

\subsection* {Acknowledgments}
The CCAT-prime project, FYST and Prime-Cam instrument have been supported by generous contributions from the Fred M. Young, Jr. Charitable Trust, Cornell University, and the Canada Foundation for Innovation and the Provinces of Ontario, Alberta, and British Columbia.
The construction of the FYST telescope was supported by the Gro{\ss}ger{\"a}te-Programm of the German Science Foundation (Deutsche Forschungsgemeinschaft, DFG) under grant INST 216/733-1 FUGG, as well as funding from Universit{\"a}t zu K{\"o}ln, Universit{\"a}t Bonn and the Max Planck Institut f{\"u}r Astrophysik, Garching.
The completion and deployment of the Prime-Cam instrument with the initial instrument modules is supported by a generous contribution from Alex Gerko, Founder and CEO of XTX Markets.
The construction of the 850 GHz instrument module for Prime-Cam is supported by CFI grants: 39656 and 46097 and Canadian provincial matching funds.
This research was supported in part by grant NSF PHY-2309135 to the Kavli Institute for Theoretical Physics (KITP) during the fall 2025 residence (SCC).
AH acknowledges support and facilities from NRC Herzberg Astronomy and Astrophysics.



\bibliography{ref}   
\bibliographystyle{spiejour}   

\end{document}